\documentclass[compactaffiliation]{Interspeech}

\interspeechcameraready

\usepackage[acronym, shortcuts, nohypertypes={acronym}]{glossaries}
\usepackage{color}
\usepackage{newtxtext}
\newacronym{AI}{AI}{Artificial Intelligence}
\newacronym{AST}{AST}{Audio Spectrogram Transformer}
\newacronym{ASR}{ASR}{Automatic Speech Recognition}
\newacronym{ASD}{ASD}{Autism Spectrum Disorder}
\newacronym{BERT}{BERT}{Bidirectional Encoder Representations from Transformers}
\newacronym{BPTT}{BPTT}{Backpropagation Through Time}
\newacronym{ComParE}{ComParE}{Computational Paralinguistics Challenge}
\newacronym{CNN}{CNN}{Convolutional Neural Network}
\newacronym{CRF}{CRF}{Conditional Random Field}
\newacronym{defcomm}{\textsc{DefComm-DB}}{Multimodal Defensive Communication Database}
\newacronym{DMRS}{DMRS}{Defense Mechanisms Rating Scale}
\newacronym{DMRS-Q}{DMRS-Q}{Q-sort based Defense Mechanisms Rating Scale}
\newacronym{egemaps}{eGeMAPS}{extended Geneva Minimalistic Acoustic Parameter Set}
\newacronym{ELECTRA}{ELECTRA}{Efficiently Learning an Encoder that Classifies Token Replacements Accurately}
\newacronym{FAU}{FAU}{Facial Action Unit}
\newacronym{GRU}{GRU}{Gated Recurrent Unit}
\newacronym{HNR}{HNR}{Harmonics to Noise Ratio}
\newacronym{LLD}{LLD}{Low-Level Descriptor}
\newacronym{LM}{LM}{Language Model}
\newacronym{LLM}{LLM}{Large Language Model}
\newacronym{LPC}{LPC}{Linear Predictive Coding}
\newacronym{LSP}{LSP}{Line Spectral Pair}
\newacronym{LSTM}{LSTM}{Long Short-Term Memory}
\newacronym{MADUV}{MADUV}{Mice Autism Detection via Ultrasound Vocalization Challenge}
\newacronym{MFA}{MFA}{Montreal Forced Aligner}
\newacronym{MFCC}{MFCC}{Mel-Frequency Cepstral Coefficient}
\newacronym{ML}{ML}{Machine Learning}
\newacronym{MLM}{MLM}{Masked Language Model}
\newacronym{openSMILE}{\textsc{openSMILE}}{Open-source Speech and Music Interpretation by Large-space Extraction}
\newacronym{OpenCV}{OpenCV}{Open Source Computer Vision}
\newacronym{PLP-CC}{PLP-CC}{Perceptual Linear Prediction and Cross-Correlation}
\newacronym{Py-Feat}{Py-Feat}{Python Facial Expression Analysis Toolbox}
\newacronym{RBF}{RBF}{Radial Basis Function}
\newacronym{RNN}{RNN}{Recurrent Neural Network}
\newacronym{SER}{SER}{Speech Emotion Recognition}
\newacronym{SSD}{SSD}{Spectral Shape Descriptor}
\newacronym{SMOTE}{SMOTE}{Synthetic Minority Over-sampling Technique}
\newacronym{SVM}{SVM}{Support Vector Machine}
\newacronym{UAR}{UAR}{Unweighted Average Recall}
\newacronym{USV}{USV}{Ultrasound Vocalization}
\newacronym{ViT}{ViT}{Vision Transformer}
\newacronym{W2V}{W2V}{wav2vec 2.0}
\newacronym{WER}{WER}{Word Error Rate}

\usepackage{hyperref}
\usepackage{cleveref}
\usepackage{xcolor}

\title{MADUV: The 1st INTERSPEECH Mice Autism Detection via Ultrasound Vocalization Challenge}

\author[affiliation={1,2}]{Zijiang}{Yang}
\author[affiliation={1}]{Meishu}{Song}
\author[affiliation={3}]{Xin}{Jing}
\author[affiliation={4}]{Haojie}{Zhang}
\author[affiliation={4}]{Kun}{Qian} 
\author[affiliation={4}]{Bin}{Hu}
\author[affiliation={5}]{Kota}{Tamada}
\author[affiliation={5}]{\\Toru}{Takumi}
\author[affiliation={3,6}]{Björn W.}{Schuller}
\author[affiliation={1}]{Yoshiharu}{Yamamoto}

\affiliation{}{The University of Tokyo}{Japan}
\affiliation{}{University of Augsburg}{Germany}
\affiliation{CHI}{TUM}{Germany}
\affiliation{}{Beijing Institute of Technology}{China}
\affiliation{}{Kobe University}{Japan}
\affiliation{GLAM}{ICL}{UK}
\email{zijiang.yang@ieee.org}


\usepackage{comment}

\begin{document}

\maketitle

\begin{abstract}
    The Mice Autism Detection via Ultrasound Vocalization (MADUV) Challenge introduces the first INTERSPEECH challenge focused on detecting Autism Spectrum Disorder (ASD) in mice through their vocalizations. Participants are tasked with developing models to automatically classify mice as either wild-type or ASD models based on recordings with a high sampling rate. Our baseline system employs a simple CNN-based model using three different features. Results demonstrate the feasibility of automated ASD detection, with the considered audible-range features achieving the best performance (UAR of $0.600$ for segment-level and $0.625$ for subject-level classification). This challenge bridges speech technology and biomedical research, offering opportunities to advance our understanding of ASD models through machine learning approaches. The findings suggest promising directions for vocalization analysis and highlight the potential value of audible and ultrasound vocalizations in ASD detection.
\end{abstract}

\section{Introduction}
\label{sec:introduction}

    \ac{ASD} is a complex neurodevelopmental condition that affects social interaction, communication, and behavior~\cite{lord2018autism}. While human studies provide critical insights, animal models, particularly mice, are essential for further understanding the genetic and neurological underpinnings of \ac{ASD}~\cite{caston1998animal, chung2004whole, nakatani2009abnormal}. According to~\cite{beaudet2007autism}, $10$--$20$\,\% of autism cases are caused by abnormal chromosomes. Several studies have sought to identify the genetic mechanisms behind at least some cases of autism~\cite{belmonte2004autism, cook2008copy}. As mice and humans share the affected genome regions, chromosomal engineering~\cite{van2006mouse} can be utilized to design mice with the same genetic defects that are proven to account for many cases of autism in humans. When undergoing behavioral tests, such mice have been shown to exhibit behavioral and affective patterns that are comparable to those seen in humans diagnosed with \ac{ASD}~\cite{nakatani2009abnormal}, making them a potential model of human \ac{ASD}. 
    
    One area that has garnered interest in this context is the analysis of \acp{USV} produced by mice~\cite{sangiamo2020ultrasonic}. These vocalizations, inaudible to humans, have been shown to vary significantly between wild-type (healthy) mice and those genetically engineered to model \ac{ASD}~\cite{nakatani2009abnormal}. In this work, we introduce the \ac{MADUV}, in which participants are tasked with building models to automatically classify mice as either \ac{ASD} or wild-type based on their \acp{USV}. The challenge makes use of data collected in the study~\cite{nakatani2009abnormal}, comprising around $7$ hours of \acp{USV} by $84$ different subjects. While numerous previous studies have analyzed human vocalizations for \ac{ASD} detection~\cite{schuller2013asc, schmitt2016towards, mohanta2022analysis}, \ac{MADUV} extends the scope to non-human subjects, making it a unique problem for computational models to solve. Given the plethora of works addressing different aspects of human speech, participants are particularly encouraged to explore the potential of transferring models and techniques on human data to ultrasonic vocalizations from mice. The baseline experiments for \ac{MADUV} as presented in this paper demonstrate the feasibility of the general approaches. The potential of leveraging methods that were proven in human speech for detecting autism in mice is indicated by the baseline results presented in this paper, highlighting the versatility of machine learning and its ability to bridge gaps between species in behavioral and neurological research. Thus, as the first animal health challenge in INTERSPEECH, \ac{MADUV} provides an opportunity to contribute to both speech technology and biomedical research by advancing our understanding of \ac{ASD} in different species through machine learning for speech data. 
    
    This challenge leverages mice models, widely used in neuroscience, to study fundamental biological mechanisms of neural development and communication in a controlled, ethically guided framework. While findings enhance our understanding of basic neural processes, they cannot be directly translated to humans. Importantly, this research focuses solely on mice vocalisations and is not intended for human screening, diagnosis, or intervention. The challenge aims to contribute to broader scientific knowledge, acknowledging the complexity of autism in humans, which cannot be fully captured by animal models.
    
    The remainder of this paper is organized as follows: \Cref{sec:related_work} reviews previous work on \ac{ASD} detection in mice. In~\Cref{sec:data}, the challenge's dataset is introduced. Our baseline experiments are outlined in~\Cref{sec:experiments}, and their results are presented in~\Cref{sec:results}. \Cref{sec:orga} provides more details on the challenge organization, before \Cref{sec:conclusion} concludes the paper.

\section{Related Work}
\label{sec:related_work}

    As for humans, there is a plethora of evidence for prosodic differences between individuals with and without ASD. For example, individuals with ASD have been found to exhibit a comparably slow speech rate~\cite{patel2020acoustic} and unusually melodic intonation~\cite{wehrle2022new}. See~\cite{asghari2021distinctive} for a recent survey on prosodic peculiarities observed in humans with ASD. 
    
    The existence of systematic patterns in ASD patients' speech motivates the application of machine learning to effectively detect ASD from human speech. Automatic autism detection from (children) speech was posed as an INTERSPEECH challenge as early as 2013~\cite{schuller2013interspeech}. Marchi~\textit{et al.}~\cite{marchi2015voice} utilize handcrafted features such as eGeMAPS~\cite{eyben2016gemaps} in combination with \acp{SVM} to distinguish ASD-affected individuals' speech from speech of individuals with typical development, achieving more than $75$\,\% \ac{UAR} on this binary classification problem. In~\cite{cho2019automatic}, both acoustic and textual features are leveraged, leading to an AUC value of around $0.75$, which, however, falls short of human performance. Investigating both English and Cantonese speakers, Lau~\textit{et al.}~\cite{lau2022cross} study cross-linguistic patterns in the speech of \ac{ASD}-affected individuals using \acp{SVM}. Their findings suggest that, in contrast to intonation-related features, rhythm-related features may be robust markers of \ac{ASD} across the two different languages. Recently, Chi~\textit{et al.}~\cite{chi2022classifying} employed random forest, \acp{CNN}, and a fine-tuned Wav2Vec 2.0 model to identify autism from self-recorded speech samples, observing that the deep learning-based methods substantially outperform the traditional random forest approach.
    
    Though mice also produce sounds in a range audible to humans, their ultrasound vocalizations are more frequent and have been proven informative with regard to a range of characteristics, including sex, age, and health conditions~\cite{yao2023review}. Ivanenko\,\textit{et al.}~\cite{ivanenko2020classifying} successfully utilized deep neural networks to determine the sex of mice based on USVs. Vogel\,\textit{et al.}~\cite{vogel2019quantifying} used supervised learning with random forests to predict $9$ expert-defined mice vocalization types, achieving $85$\,\% classification accuracy. An alternative to expert-defined categories, Wang\,\textit{et al.}~\cite{wang2020bringing} propose an unsupervised spectral clustering approach to identify mice~\ac{USV} types. In the most similar work to \ac{MADUV}, Qian\,\textit{et al.}~\cite{qian2021sensing} conducted a pilot study using a large-scale pre-trained audio neural network to extract high-level features from mice \acp{USV} for identifying ASD model mice. Their approach achieved a \ac{UAR} of $66.6$\,\% in this binary classification task, marking the first use of inaudible \acp{USV} for detecting \ac{ASD} type in mice.

\section{Data}
\label{sec:data}

    The dataset contains recordings of $84$ subjects, of which $44$ ($30$ male, $14$ female) belong to the wild-type and $40$ ($27$ male, $13$ female) are~\ac{ASD} model types. For all subjects, we collect one recording at an early development stage, more specifically $8$ days after their birth, i.e., Postnatal Day $8$ (P$08$). During this early developmental stage, mice emit an increased amount of~\ac{USV}s due to anxiety-like behavior induced by separating them from their mothers~\cite{nakatani2009abnormal}. Each recording is around $5$ minutes long, resulting in about $7$ hours of audio in total. All recordings are obtained via high-precision microphones (Avisoft UltraSoundGate 416H) with a sample rate of $300$\,kHz.
    
    The dataset underwent stratified partitioning to ensure subject independence while maintaining balanced distributions of sex and ASD model type. For evaluation purposes, approximately $20$\,\% of subjects were allocated to the test set first, consisting of $12$ male and $4$ female subjects. To enhance the statistical robustness of the evaluation, each test sample was segmented into non-overlapping $30$-second clips, yielding $160$ audio samples. In accordance with the challenge setup, subject IDs and their corresponding labels were withheld from the test set.
    
    Subsequently, for the baseline system implementation, the remaining subjects were partitioned into training and validation sets, with $51$ subjects (approximately $60$\,\%) allocated to the training set and $17$ subjects (approximately $20$\,\%) assigned to the validation set. The distribution of the ASD model type maintained consistency across partitions, comprising $47.06$\,\% ($24$\,/\,$51$) of training subjects, $47.06$\,\% ($8$\,/\,$17$) of validation subjects, and $50$\,\% ($8$\,/\,$16$) of test subjects. Gender distribution exhibited a male predominance across all partitions, with males constituting $64.71$\,\% ($33$\,/\,$51$) of the training set, $70.59$\,\% ($12$\,/\,$17$) of the validation set, and $75$\,\% ($12$\,/\,$16$) of the test set.
    
    For the training and validation sets, participants will have access to the complete $5$-minute audio recordings, allowing them free implementation of customized segmentation approaches and training-validation partition strategies. As mentioned above, the test set recordings underwent systematic segmentation into $160$ $30$-second clips. To maintain assessment integrity, all labels of the test set were removed, and the clips were systematically shuffled according to a randomly generated rule that will be used in the evaluation. The partition statistics are detailed in~\Cref{tab:statistics}.

    \begin{table}[t!]
    \centering
    \resizebox{\columnwidth}{!}{
    \begin{tabular}{lrrrr}
        \hline
        & \multicolumn{1}{c}{\textbf{train}} & \multicolumn{1}{c}{\textbf{valid}} & \multicolumn{1}{c}{\textbf{test}} & \multicolumn{1}{c}{\textbf{total}} \\ \hline
        \# subjects & 51 & 17 & 16 & 84 \\
        ~~~of which \ac{ASD} & 24 & 8 & 8 & 40 \\
        male\,/\,female & 33\,/\,18 & 12\,/\,5 & 12\,/\,4 & 57\,/\,27 \\
        ~~~of which \ac{ASD} & 15\,/\,9 & 6\,/\,2 & 6\,/\,2 & 27\,/\,13 \\
        duration & 4:15:10 & 1:25:04 & 1:19:59 & 7:00:14 \\ \hline
    \end{tabular}}
    \caption{Dataset statistics for the entire dataset (\textbf{total}) and the training (\textbf{train}), validation (\textbf{valid}), and \textbf{test} partitions. Durations are given as \textit{hours\,:\,minutes\,:\,seconds}.}\label{tab:statistics}
    \end{table}

    It should be clarified that this challenge uses mice models to study basic neural development mechanisms, adhering to ethical standards for animal research. The research focuses solely on analysing mice vocalisations and does not aim for human diagnosis or intervention. The challenge contributes to neuroscience by advancing understanding of neural development, while recognising the limitations of animal models in representing human conditions like autism.

\section{Baseline Experiments}
\label{sec:experiments}

    To address the constraints of limited training and validation data, the $5$-minute recordings in both subsets were systematically segmented into $30$-second clips with $15$-second overlap. Clips of insufficient duration were excluded from the dataset due to their limited information. This segmentation approach yielded $19$ clips per original sample, resulting in a total of $969$ training samples and $323$ validation samples.
    
    In order to establish a simple, yet robust benchmark, we train a \ac{CNN}-based model employing different feature sets. We select three different spectrogram-based feature sets widely applied for human speech research, namely \textit{full}, \textit{ultra}, and \textit{audi} as will be detailed.
    
    Spectrograms and their variant Mel-spectrograms represent a well-established and conventional approach to feature extraction in audio-based machine learning applications. These time-frequency/quefrency representations have demonstrated robust performance across diverse domains, including acoustic scene classification~\cite{zhang2021acoustic, ren2018deep}, animal vocalization classification~\cite{pahuja2021sound, nanni2020spectrogram}, and speech emotion recognition~\cite{ottl2020group, zhao2019exploring}. Consequently, spectrogram-based feature extraction presents a methodologically sound basis for the baseline system implementation.
    
    The feature extraction methods began with a unified preprocessing approach for all three feature types, beginning with spectrogram generation at a sampling rate of $300$\,kHz by using~\texttt{Scipy}\footnote{https://scipy.org/}. To optimize frequency resolution, both the window size and the number for the Fast Fourier Transform (FFT) were configured at $300,\,000$. The hop length was set to $150,\,000$, resulting in a temporal dimension of $59$ frames.
    
    Subsequently, the spectrogram was partitioned into distinct frequency ranges. The ultrasonic component, encompassing frequencies between $20$\,kHz and $150$\,kHz, was extracted to constitute the \textit{ultra} feature set. Correspondingly, the audible spectrum below $20$\,kHz was separated to form the \textit{audi} feature set. The \textit{full} feature set, as its name suggests, retained the complete spectral information across all frequency ranges.
    
    To mitigate the computational expense and optimize processing efficiency, dimensionality reduction was implemented by applying average bins. Specifically, the frequency dimension was standardized to $500$ bins across all feature sets through the application of different sizes of average bins. The feature sets maintained uniform dimensions of $[59,500]$. This feature extraction methodology provides a robust framework for the baseline experiments while facilitating the investigation of frequency-band-specific information.
    
\begin{figure}[t!]
        \centering
        \includegraphics[scale=0.75]{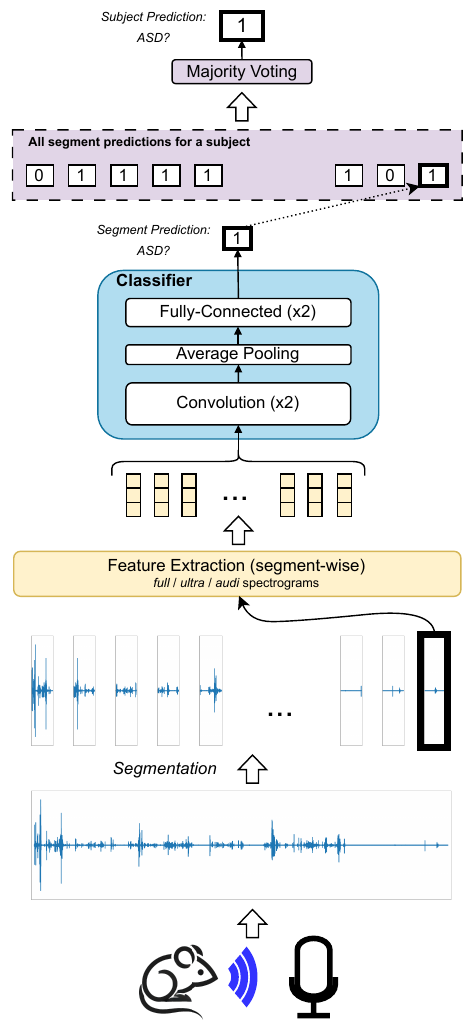}
        \caption{End-to-end overview of the baseline approach for one subject. More specifically, the figure shows the prediction of the last of the subject's segments, followed by the majority voting to predict \ac{ASD} for the subject.}
        \label{fig:system}
    \end{figure}

    As a simple, yet robust benchmark system, we employ a~\ac{CNN} model for our baseline experiments. The architecture begins with two stacked convolutional layers, followed by two fully connected layers. \Cref{fig:system} shows an overview of the entire approach. The model is trained using the Adam optimizer and the binary cross-entropy loss function. The output logits undergo \textit{sigmoid} activation to generate probabilistic predictions, with the decision threshold for binary classification treated as a hyperparameter. The optimal threshold value is determined through a grid search, from $0.10$ to $0.90$ with an interval of $0.05$, during the training phase. After the model achieved the best performance on the validation set, it was utilized subsequent predictions on the test set. All baseline experiments were conducted using~\texttt{PyTorch}\footnote{https://pytorch.org/}.
    
    \ac{MADUV} introduces two evaluation criteria: segment-level and subject-level classification. During both the training and testing phases, the trained model predicts the class of all segments in the validation and test sets, respectively. To determine the subject-level classification, majority voting is applied to the segment predictions belonging to the same subject. For instance, if a greater number of segments for a subject are classified as wild-type rather than~\ac{ASD}, the subject is categorized as wild-type.
    
    The baseline experiments are designed to mimic the challenge setup where participants can submit up to $5$ predictions (cf.~\Cref{sec:orga}). Therefore, for each feature set, we train the model $5$ times using $5$ different random seeds.

\section{Results}
\label{sec:results}

    Given the significance of both ASD and wild-type classes, the \ac{UAR}--accounting for both classes in the dataset--is selected as the evaluation criterion for the challenge. As mentioned above, we compute UAR for both predictions in segment-level and subject-level evaluation.
    
    \begin{table*}[t!]
    \centering
    \resizebox{0.9\textwidth}{!}{
    \begin{tabular}{lrrrr}
        \hline
        & \multicolumn{4}{c}{[UAR$\uparrow$]} \\ & \multicolumn{2}{c}{\textbf{Validation Set}} & \multicolumn{2}{c}{\textbf{Test Set}} \\ 
        \textbf{Feature}   & \multicolumn{1}{c}{\textbf{Segment}} & \multicolumn{1}{c}{\textbf{Subject}} & \multicolumn{1}{c}{\textbf{Segment}} & \multicolumn{1}{c}{\textbf{Subject}} \\ \hline
        \textbf{\textit{full}} & .675 (.666\,$\pm$\,.007) & .813 (.777\,$\pm$\,.033) & .581 (.556\,$\pm$\,.029) & .625 (.575\,$\pm$\,.028) \\
        \textbf{\textit{ultra}} & .664 (.649\,$\pm$\,.017) & .819 (.760\,$\pm$\,.040) & .569 (.530\,$\pm$\,.027) & .625 (.562\,$\pm$\,.063) \\ 
        \textbf{\textit{audi}} & .682 (.674\,$\pm$\,.009) & .813 (.763\,$\pm$\,.052) & \textbf{.600} (.588\,$\pm$\,.016) & \textbf{.625} (.588\,$\pm$\,.034) \\
        \hline
    \end{tabular}}
    \caption{UAR of baseline experiments' results. For each feature, we report results at the segment-level and subject-level for both validation and test set. The first value in the \textit{Validation Set} columns is the best UAR obtained on the validation set across $5$ fixed seeds. The first value in the \textit{Test Set} columns gives the result on the test set from the run that led to the best validation performance. The values in brackets provide mean and standard deviations across the $5$ seeds for both the validation and test data.}\label{tab:results}
    \end{table*}
    
    \Cref{tab:results} presents the results. Consistent with the challenge setup, we not only just report means and standard deviations across the $5$ seeds, but also the best result in terms of performance on the validation set and the corresponding model's result on the test set. 
    
    The results demonstrate the effectiveness of different feature extraction methods in addressing both segment-level and subject-level classification tasks, highlighting promising directions for further exploration. Analyses reveal comparable performance metrics across all three feature sets on the validation set, suggesting effective feature learning capabilities of the~\ac{CNN} architectures during the training phase. Due to the implementation of majority voting, a direct proportional correlation between subject-level and segment-level UAR is not observed. Furthermore, subject-level evaluation metrics are derived from optimal segment-level performance. When considering segment-level metrics exclusively, the \textit{audi} feature set demonstrates superior performance on the validation set, achieving maximum and mean UAR values of $0.682$ and $0.674$, respectively.
    
    The \textit{audi} feature set demonstrates robust predictive performance on the test set. Across five random seeds, it achieves superior segment-level performance with a maximum UAR of $0.600$ and mean UAR of $0.588$, surpassing the other feature sets. While all three feature sets achieve equivalent maximum subject-level UAR values of $0.625$, the $audi$ feature set maintains superiority in mean performance metrics. Consequently, the segment-level UAR of $0.600$ and subject-level UAR of $0.625$ are established as the benchmark for the challenge.
    
    Besides, statistical significance analysis was conducted on the baseline results utilizing a one-tailed one-sample t-tests across five random seeds. Given the binary classification task, the theoretical chance-level UAR is $0.500$ for both segment and subject-level analyses. The results demonstrate that even the least performant feature set (\textit{ultra}) achieved statistically significant improvement over the chance level at $p<0.05$. The superior \textit{audi} feature set exhibited highly significant performance gains relative to chance level, with $p<0.0005$ and $p<0.005$ for segment-level and subject-level analyses, respectively. Furthermore, comparative analysis between feature sets revealed the \textit{audi} feature's significant superiority over the \textit{ultra} feature set ($p<0.001$).
    
    While Nakatani~\textit{et al.}~\cite{nakatani2009abnormal} demonstrated differential USVs frequencies between wild-type and ASD model mice over $5$-minute recordings, our baseline findings indicate superior utility of audible-range vocalizations compared to USVs. This apparent discrepancy can be reconciled by considering distinct underlying mechanisms: while separation anxiety causes increased ultrasonic calling behavior, the spectral characteristics of audible vocalizations appear to contain more discriminative features of ASD type that are beneficial to neural network-based classification. This novel finding contributes meaningfully to our understanding of vocalizations in ASD research.
    
    To summarise, the findings demonstrate the efficacy of spectrogram-based feature extraction methodologies, with audible-range spectrograms exhibiting superior capacity for ASD classification at both segment and subject levels. This baseline implementation establishes a foundation for future methodological innovations, including alternative data partitioning strategies, preprocessing schemes, feature engineering approaches, and neural network architectures. Moreover, the robust performance of audible-range vocalizations suggests a potential similarity between mice and human vocalizations of the ASD model, and it is worth further investigation of cross-species behavioral models.

\section{Challenge Organization}
\label{sec:orga}

    In order to participate in \ac{MADUV}, teams must register under the lead of a professor in academia, or a research team leader in industry. Upon signing the EULA, they obtain access to the dataset, the baseline code as well as the evaluation system. Teams are allowed to comprise at most $5$ members excluding the PI. Moreover, we do not allow one and the same person to be part of several teams. None of the organizers will be members of any participating team.
    
    The evaluation system will be hosted on EvalAI\footnote{https://eval.ai/web/challenges/challenge-page/2428/}, an established platform for shared task evaluation. This way, there will be a transparent leaderboard and automatic enforcement of the constraints on the number of submissions. Participants submit their segment-level predictions for the test set and immediately receive their results. Note that the test subject IDs are not disclosed to the participants. The subject-level results are calculated using the same majority voting approach above.
    
    Analogously to the baseline evaluation (cf.~\Cref{tab:results}), we evaluate participants' submissions on both the segment and the subject level via the respective UARs. This leads to two rankings, which are averaged to obtain an overall rank. In case the two teams' average ranks are equal, priority is given to the team with a lower rank on the segment level. Officially winning \ac{MADUV} requires to submit a system paper describing the technical approach, which must also be accepted into the conference.
    
    In order to facilitate the reproducibility of the results reported above and enable participants to rapidly get started with developing their approaches, we make the code for our baseline systems publicly available\footnote{https://github.com/KamijouMikoto/maduv\_2025/}. Furthermore, participants can download all $5$ pre-extracted feature sets and the relevant checkpoints of our baseline models. All up-to-date information on the challenge logistics can be found at the MADUV website\footnote{https://www.maduv.org/}.

\section{Conclusion}
\label{sec:conclusion}

    In this baseline paper, we introduced \ac{MADUV}, the 1st INTERSPEECH Mice Autism Detection via Ultrasound Vocalization Challenge. We described the extensive challenge dataset and reported a set of competitive baseline results that serve as benchmarks for participants' approaches. While all feature sets considered lead to above-chance performance, the results obtained with the audible spectrogram prove to be particularly encouraging. These insights not only enhance our understanding of mice vocalizations but also pave the way for leveraging human speech technologies to better analyze animal communication and behaviors. Besides those aiming at the merely quantitative task of outperforming the baselines, we also look forward to contributions that deal with the dataset in a more qualitative manner and thus increase our understanding of the data.

\section{Acknowledgements}
\label{sec:acknowledgements}

    The paper is supported by the Japan Science and Technology (JST) Agency MOONSHOT R\&D (Grant JPMJMS229B and JPMJMS2021). Furthermore, this paper is supported by MDSI -- the Munich Data Science Institute as well as MCML -- the Munich Center of Machine Learning. Bj\"orn W. Schuller is also with the Konrad Zuse School of Excellence in Reliable AI (relAI), Munich, Germany. 

\newpage
\bibliographystyle{IEEEtran}
\bibliography{mybib}

\end{document}